\documentclass{article} 
\usepackage{iclr2022_conference,times}

\usepackage{amsmath,amsfonts,bm}

\def\eqref#1{equation~\ref{#1}}

\def\1{\bm{1}}

\DeclareMathAlphabet{\mathsfit}{\encodingdefault}{\sfdefault}{m}{sl}
\SetMathAlphabet{\mathsfit}{bold}{\encodingdefault}{\sfdefault}{bx}{n}

\usepackage{hyperref}
\usepackage{url}
\usepackage{graphicx}
\usepackage{amsmath}
\usepackage{amssymb}
\usepackage{array, caption, floatrow, tabularx, makecell, booktabs}
\usepackage{caption}
\title{GCNScheduler: Scheduling Distributed Computing Applications using Graph 
Convolutional Networks}

\author{Mehrdad Kiamari and Bhaskar Krishnamachari \\
Electrical and Computer Engineering Department,\\ 
Viterbi School of Engineering,\\ 
University of Southern California, Los Angeles, USA\\
\texttt{\{kiamari,bkrishna\}@usc.edu}
}

\iclrfinalcopy 
\begin{document}

\maketitle

\begin{abstract}
We consider the classical problem of scheduling task graphs corresponding to complex applications on distributed computing systems. A number of heuristics have been previously proposed to optimize task scheduling with respect to metrics such as makespan and throughput. However, they tend to be slow to run, particularly for larger problem instances, limiting their applicability in more dynamic systems. Motivated by the goal of solving these problems more rapidly, we propose, for the first time, a graph convolutional network-based scheduler (GCNScheduler). By carefully integrating an inter-task data dependency structure with network settings into an input graph and feeding it to an appropriate GCN, the GCNScheduler can efficiently schedule tasks of complex applications for a given objective.
We evaluate our scheme with baselines through simulations. We show that not only can our scheme quickly and efficiently learn from existing scheduling schemes, but also it can easily be applied to large-scale settings where current scheduling schemes fail to handle. We show that it achieves better makespan than the classic HEFT algorithm, and almost the same throughput as throughput-oriented HEFT (TP-HEFT), while providing several orders of magnitude faster scheduling times in both cases. For example, for makespan minimization, GCNScheduler schedules 50-node task graphs in about 4 milliseconds while HEFT takes more than 1500 seconds; and for throughput maximization, GCNScheduler schedules 100-node task graphs in about 3.3 milliseconds, compared to about 6.9 seconds for TP-HEFT.
\end{abstract}

\section{Introduction}
\label{sec:introduction}

Successfully running complex graph-based applications, ranging from edge-cloud processing in IoT systems \cite{scheduling_IoT_space}-\cite{scheduling_IoT_software} to processing astronomical observations \cite{astronomical}, heavily relies on executing all sub-components of such applications through an efficient task-scheduling. Not only does    
efficient task scheduling play a crucial role in improving the utilization of computing resources and reducing the required time to executing tasks, it can also lead to significant profits to service providers \cite{profit_reduction}. In this framework, any application consists of multiple tasks with a given inter-task data dependency structure, i.e. each task generates inputs for certain other tasks. Such dependencies can be expressed via a directed acyclic graph (DAG), also known as \emph{task graph}, where vertices and edges represent tasks and inter-task data dependencies, respectively. An input job for an application is completed once all the tasks are executed by compute machines according to the inter-task dependencies. 

There are two commonly used metrics for schedulers to optimize: makespan and throughput. The required time to complete all tasks for a single input is called the makespan.  The maximum steady state rate at which inputs can be processed in a pipelined manner is called throughput. Makespan minimization and throughput maximization can each be achieved through relevant efficient \emph{task-scheduling} algorithms that assign tasks to appropriate distributed computing resources to be executed.

The underlying methodology for task scheduling can be categorized into heuristic-based (e.g. \cite{heuristic_new_Eskandari}- \cite{new_heuristic_journal_Pham}), meta-heuristic ones
(e.g. \cite{Kennedy-optimization},\cite{ADDYA-Simulatedannealing}-\cite{Fan-Simulated-Annealing}, \cite{SHISHID-Genetic-based},\cite{DASGUPTA-Genetic},\cite{genetic_Izadkhah}), and optimization-based schemes (e.g. \cite{Azar-convex-unrelated}-\cite{Skutella-SDP-no-comm}). 

One of the most well-known heuristic scheduling schemes for makespan minimization is the heterogeneous earliest-finish time (HEFT) algorithm~\cite{HEFT}, which will be considered as one of our benchmarks. For throughput maximization, we benchmark against the algorithm presented in~\cite{HEFT-TP} which we refer to as TP-HEFT. 

One of the fundamental disadvantages of all the above-mentioned scheduling schemes is that they work well only in relatively small settings; once a task graph becomes large or extremely large, they require very long computation times. We anticipate that applications in many domains, such as IoT for smart cities will result in increasingly complex applications with numerous inter-dependent tasks, and scheduling may need to be repeated quite frequently in the presence of network or resource dynamics~\cite{dustdar2017smart, syed2021iot}. Therefore, it is essential to design a faster method to schedule tasks for such large-scale task graphs. 

A promising alternative is to apply machine learning techniques for function approximation to this problem, leveraging the fact that scheduling essentially has to do with finding a function mapping tasks to compute machines. Given the graph structure of applications, we propose to use an appropriate 
graph convolutional network (GCN) \cite{kipf} to schedule tasks through learning the inter-task dependencies of the task graph as well as network settings (i.e., execution speed of compute machines and  communication bandwidth across machines) in order to extract the relationship between different entities. The GCN has attracted significant attention in the literature for its ability in addressing many graph-based applications to perform semi-supervised link prediction \cite{link_pred} and node classification \cite{kipf}. The idea behind GCN is to  construct node embeddings layer by layer. In each layer, a node embedding is achieved by aggregating its neighbors' embeddings, followed by a neural network (i.e. a linear transformations and nonlinear activation). 
In case of node classification, the last layer embedding is given to a softmax operator to predict node labels, and consequently the parameters of GCN can be learned in an end-to-end manner. In general, there are two types of GCNs, namely spectral-based GCNs~\cite{spectralGCN} and special-based ones~\cite{graphsage}. To obtain node embedding, the former requires matrix decomposition of Laplacian matrix (which results in scalability issues due to non-linear computation complexity of the decomposition) while the latter does not have such complexity thanks to the idea of message-passing. 

To the best of our knowledge, there is no prior work that has proposed a pure \emph{spatial-based} GCN, incorporated with carefully-designed the features of both nodes and edges for task graphs, to perform scheduling over distributed computing systems. 

The main contributions of this paper are as follows:
\begin{itemize}
    \item We propose GCNScheduler, which can quickly schedules tasks by carefully integrating a task graph with network settings into a single input graph and feeding it to an appropriate GCN model.
    
    \item Any existing scheduling algorithm can be used as a teacher to train GCNScheduler, for any metric. We illustrate this by training GCNScheduler using HEFT~\cite{HEFT} for makespan minimization, and TP-HEFT~\cite{HEFT-TP} for throughput maximization.

    \item We evaluate the performance of our proposed scheme and show that, not only can our GCNScheduler be trained in a very short period of time\footnote{For instance, it takes around $<$15 seconds to train a graph with 8,000 nodes.}, it also gives scheduling performance comparable to the teacher  algorithm. We show our approach gives comparable or better scheduling performance in terms of makespan with respect to HEFT and throughput with respect to TP-HEFT, respectively.
    
    \item We show that GCNScheduler is several orders of magnitude faster than previous heuristic algorithms in obtaining the schedule for a given task graph. For example, for makespan minimization, GCNScheduler schedules 50-node task graphs in about 4 milliseconds while HEFT takes more than 1500 seconds; and for throughput maximization, GCNScheduler schedules 100-node task graphs in about 3.3 milliseconds, compared to about 6.9 seconds for TP-HEFT.
    
    \item We show that GCNScheduler is able to efficiently perform scheduling for \emph{any size} task graph. In particular, we show that our proposed scheme is able to operate over large-scale task graphs where existing schemes require excessive computational resources.
    
\end{itemize}

\subsection{Related Work}\label{related_work}

Task scheduling can be categorized into multiple groups from different perspectives. 
One way of categorizing task scheduling schemes has to do with the type of algorithms used for assigning tasks to compute resources. Heuristic, meta-heuristic, and optimization-based are three categories of task scheduling schemes. Heuristic task scheduling schemes can be divided into quite a few sub categories based on their approach, such as load balancing \cite{Ren-loadbalance-2012},\cite{Bhatia-loadbalance},\cite{KUMAR-loadbalance}, priority-based scheduling~\cite{HEFT},\cite{SUDARSAN-priority},\cite{Dubey-priority},\cite{list_scheduling}, task  duplication \cite{duplication}, and clustering \cite{clustering}. 

Since heuristic algorithms may sometimes perform poorly compared to optimal task scheduling, meta-heuristic (e.g. Particle Swarm Optimization \cite{Kennedy-optimization}, Simulated Annealing \cite{ADDYA-Simulatedannealing}-\cite{Fan-Simulated-Annealing}, Genetic-based approach \cite{SHISHID-Genetic-based},\cite{DASGUPTA-Genetic},\cite{genetic_Izadkhah}) and optimization-based schemes (e.g. \cite{Azar-convex-unrelated}-\cite{Skutella-SDP-no-comm}), which aim at approximating the NP-hard optimization of task scheduling, have attracted significant attention. 
However, all the above-mentioned heuristic, meta-heuristic and optimization-based schemes tend to run extremely slowly as number of tasks becomes large due to iterative nature of these methods, which requires excessive computations. Moreover, this issue makes the aforementioned schemes unable to handle large-scale task graphs.

As obtaining the optimal scheduler is basically the same as finding an appropriate mapper function,  which maps tasks to compute machines, machine-learning based scheduling has begun emerging as an alternative thanks to advances in fundamental learning methods, such as deep learning~\cite{deeplearning} and reinforcement learning (RL)~\cite{reinforcementlearning}. 
Sun \emph{et. al.} proposed DeepWave~\cite{Deepwave-sun}, a scheduler which reduces job completion time using RL while specifying a priority list \footnote{Which indicates the scheduling priority of edges in a job DAG.} as the action and the completion time of a job DAG as the reward.
Furthermore, Decima~\cite{Decima} schedules tasks over a spark cluster by training a neural network using RL with scheduling the next task for execution as the action and a high-level scheduling objective of minimizing the makespan as the reward.  The aforementioned RL-based schemes suffer from having a huge action space (i.e., the space of scheduling decisions).

While Decima~\cite{Decima} only operates in homogeneous environment, Luo \emph{et. al.} proposed Lachesis~\cite {Lachesis} to operate over heterogeneous environment. Lachesis combines three different components, a GCN, an RL policy network, and a heuristic task mapper. There are three main differences from our work with respect to their use of GCN: first, they use the GCN to embed task nodes only without taking network settings into account as we do; second, they use a regular GCN which does not explicitly account for directed nodes while we use an EDGNN~\cite{edgnn} which does; and finally, the GCN in Lachesis does not do scheduling (only task node embedding), whereas we are the first to propose to use a GCN directly for task scheduling.


 

The remaining of the paper is organized as follows: In the next section, we elaborate upon the problem formulation. In section \ref{proposed}, we overview GCNs and explain in detail on how our proposed scheme works. Finally, in section \ref{results}, we show the numerical results on the performance of our proposed scheme against well-known approaches.

\section{Problem Statements}\label{formulation}

We now elaborate upon formally representing the minimization of makespan and the maximization of throughput as optimization problems. Every application/job is comprised of inter-task data dependencies. In order to finish a job, all its tasks require to be executed at least on a single compute machine. As far as compute machines are concerned, they are interconnected via communication links. 

Before expressing the definition of makespan and throughput, let us explain about task dependencies, referred to as \emph{task graph}, and network settings.  

\begin{figure}[h]
\includegraphics[scale=.5]{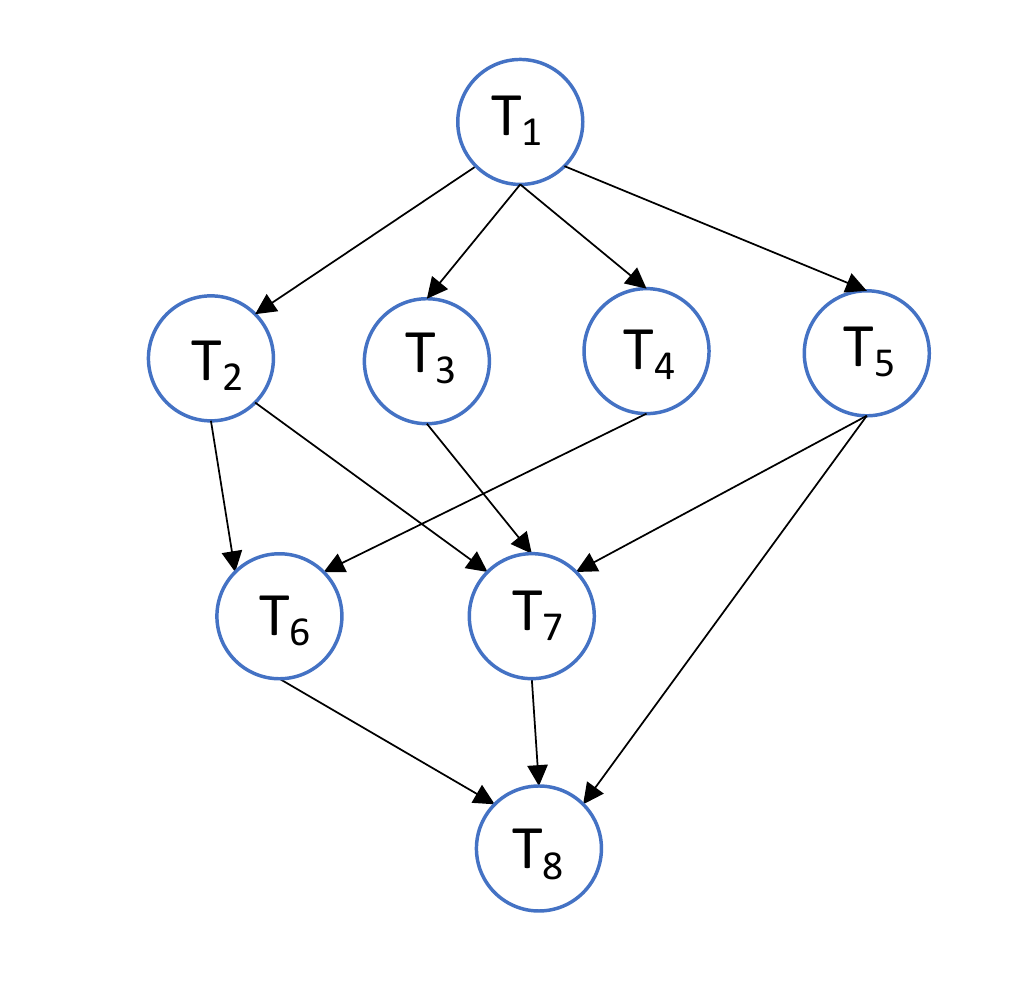}
\caption{An example of task graph, which is in the form of a DAG, with eight tasks. For instance, task $T_7$ requires tasks $T_2$, $T_3$, and $T_5$ to be first executed and generate their outputs before executing task $T_7$.}
\label{fig:task_graph}
\end{figure}

\noindent{\bf Task Graph:} Since there are dependencies across different tasks, meaning that a task generates inputs for certain other tasks, we can model this dependency through a \emph{DAG} as depicted in Fig. \ref{fig:task_graph}.  
Suppose we have $N_T$ tasks $\{T_i\}_{i=1}^{N_T}$ with a given task graph $G_{Task}:=(V_{Task},E_{Task})$ where $V_{Task}:=\{T_i\}_{i=1}^{N_T}$ and $E_{Task}:=\{e_{i,i'}\}_{(i,i')\in \Omega}$ respectively represent the set of vertices and edges (task dependencies) with $\Omega :=\{(i,i')|$ if task $T_i$ generates inputs for task $T_{i'}$ $\}$. Let us define vector ${\bf p}:= [p_1,\dots,p_{N_T}]^T$ as the amount of computations required by tasks. For every tasks $T_i$ and $T_j$, $\forall i,j$ where $e_{i,j}\in E_{Task}$, task $T_i$ produces $d_{i,j}$ amount of data for task $T_j$ after being executed by a machine.  


\noindent{\bf Network Settings:} Each task is required to be executed on a \emph{compute} node (machine) which is connected to other compute nodes (machines) through communication links (compute node and machine are interchangeably used in this paper). Let us suppose to have $N_C$ compute nodes $\{C_j\}_{j=1}^{N_C}$. Regarding the execution speed of compute nodes, we consider vector ${\bf e}:= [e_1,\dots,e_{N_C}]^T$ as the executing speed of machines. 
The communication link delay between any two compute nodes can be characterized by bandwidth. 
Let us denote $B_{i,j}$ as the communication bandwidth of the link from compute node $C_j$ to compute node $C_{j}$.
In case of two machines not being connected to each other, we can assume the corresponding bandwidth is zero (infinite time for communication delay). 

In general, a task-scheduling scheme maps tasks to compute nodes according to a given objective. Formally speaking, a task scheduler can be represented as a function $m(.):V_{Task} {\rightarrow} \{C_k\}_{k=1}^{N_C}$ where task $T_i$, $\forall i$, is assigned to machine $m(i)$. We next present two well-known objectives, namely the makespan-minimization and the throughput-maximization. 

\textbf{Objective 1:} The first objective function for the task assignment we consider is the makespan-minimization. In particular, we need to find a scheduler that assigns tasks to compute machines such that the resulting makespan is minimized. Our proposed scheme aims at obtaining such scheduler by utilizing a carefully-designed GCN where it is able to classify tasks into machines\footnote{Each machine represents a class in our problem.}.

Before formally defining the \emph{makespan}, we need to define \emph{Earliest Start Time (EST)}, \emph{Earliest Finish Time (EFT)},  \emph{Actual Start Time (AST)}, and \emph{Actual Finish Time (AFT)} as follows:

\noindent{\textit{Definition 1}}: $EST(T_i,C_j)$ denotes the earliest execution start time for task $T_i$ being executed on compute node $C_j$. Note that $EST(T_0,C_j)=0,~\forall j$.

\noindent{\textit{Definition 2}}: $EFT(T_i,C_j)$ denotes the earliest execution finish time for task $T_i$ being executed on compute node $C_j$. 

\noindent{\textit{Definition 3}}: $AST(T_i)$ and $AFT(T_i)$ denote the actual start time and the actual finish time of task $T_i$. 

Regarding the computations of the aforementioned definitions for each task, one can recursively compute them starting from task $T_1$ according to the following formula~\cite{HEFT}:
\begin{equation}\label{EST}
\begin{aligned} 
EST&(T_i,C_j)=\max \{ avail[j],
\max_{T_k:e_{k,i}\in E_{Task}}{\Big(AFT(T_k)+comm_{k,i}}\Big) \},
\\EFT&(T_i,C_j)=EST(T_i,C_j)+\frac{p_i}{e_j},
\end{aligned}
\end{equation}
where $comm_{i,j}:=\frac{data_{i,j}}{B_{m(T_i),m(T_j)}}$ and $avail[j]$ indicate the earliest time at which compute node $C_j$ is ready to execute a task.

\noindent{\textit{Definition 4 (Makespan)}}: After all tasks are assigned to compute nodes for execution, the actual time for completion of a job is equal to the actual finish time of the last task. Therefore, the makespan can be represented as 
\begin{equation}\label{makespan}
\begin{aligned} 
makespan=\max\{AFT(T_{N_T})\}.
\end{aligned}
\end{equation}

\textbf{Objective 2:} The second objective function that we consider for task-scheduling is the throughput maximization. Unlike makespan which is the overall execution time for a given input, the
throughput stands for the average number of inputs that can be executed per unit-time in steady state.
By assuming the number of inputs to be infinite and denoting $N(t)$ as the number of inputs completely executed by a scheduler at time $t$, the throughput would be $\lim_{t\rightarrow \infty}{\frac{N(t)}{t}}$. In \cite{HEFT-TP}, authors showed that the following definition characterize the throughput of a scheduler.

\noindent{\textit{Definition 5 (Throughput~\cite{HEFT-TP})}}: 
For a given task-assignment, the throughput of a scheduler is ${1}/{\tau}$ where $\tau$ is the time taken by any resource to execute an input, and it can be written as 
\begin{equation}\label{tau_def}
\begin{aligned} 
\tau:=\max \{\max_{C_q}\{t_q^{comp},t_q^{out},t_q^{in}\},\max_{C_q\rightarrow C_r}{t_{q,r}}\},
\end{aligned}
\end{equation}
with
\begin{itemize}
    \item $t_q^{comp}$: representing the computation time of compute machine $C_q$ for a single input (i.e. $t_q^{comp}=\frac{\sum_{i:m(T_i)=C_q}{p_i}}{e_q}$),  
    \item $t_q^{out}$: representing the time taken by compute machine $C_q$ for outgoing interface (i.e. $t_q^{out}=\frac{\sum_{r}{d_{q,r}^{(node)}}}{B_q^{out}}$ where $d_{q,r}^{(node)}$ and $B_q^{out}$ respectively indicate amount of data transferred from compute machine $C_q$ to $C_r$ and maximum outgoing bandwidth of compute machine $C_q$),
    \item $t_q^{in}$: representing the time taken by compute machine $C_q$ for incoming interface (i.e. 
    $t_q^{in}=\frac{\sum_{r}{d_{r,q}^{(node)}}}{B_q^{in}}$ where $B_q^{in}$ indicates the maximum incoming bandwidth of compute machine $C_q$),
    \item $t_{q,r}$: representing the communication time taken to transfer data from compute machine $C_q$ to compute machine $C_r$ (i.e. $t_{q,r}={d_{q,r}^{(node)}}/{B_{q,r}}$). 
\end{itemize}

{\noindent{\bf Remark}}: other objectives could also be considered in the future.

\section{Proposed GCNScheduler}\label{proposed}
We present a novel machine-learning based task scheduler which can be trained with respect to aforementioned objectives. Since the nature of task-scheduling problem has to do with graphs (i.e. task graph), it is essential to utilize a machine-learning approach designed for capturing the underlying graph-based relationships of the task-scheduling problem. To do so, we employ a suitable GCN\cite{edgnn}, in order to obtain a model which can automatically assign tasks to compute machines. We need to point out there is no prior GCN-based scheduling scheme incorporated with both carefully-designed features of nodes and edges. 
This novel idea has significant advantages over the conventional scheduling schemes. First, it can remarkably reduce the computational complexity compared to previous scheduling algorithms. Second, after training an appropriate GCN, our scheme can handle any large-scale task graph while conventional schemes severely suffer from the scalability issue. 


\subsection{Overview of Graph Neural Networks}
\begin{figure*}[t]
\centering
\includegraphics[scale=.45,trim= 0mm 60mm 80mm 15mm,clip=true]{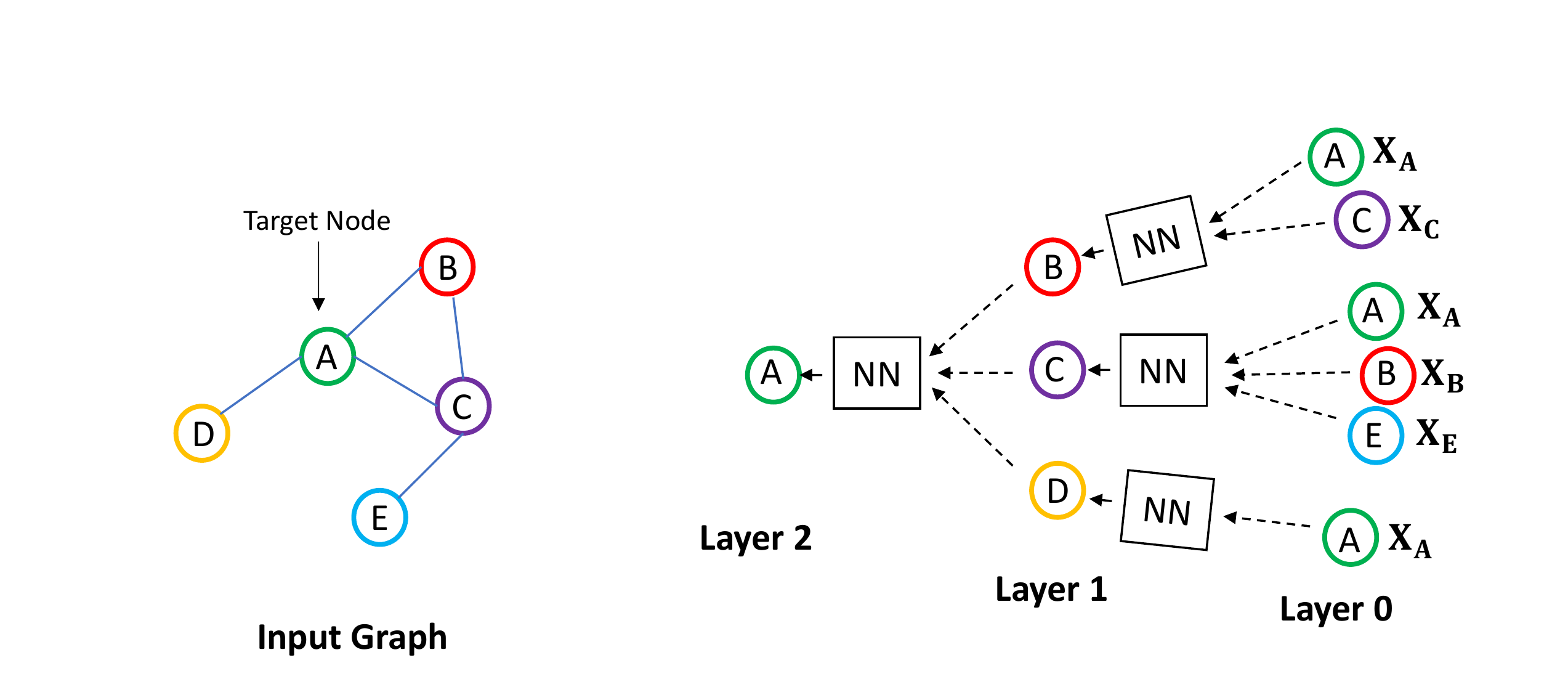}
\caption{An illustration of message-passing for the input graph shown on the left side. Regarding the notation, $X_A$ represents node A's feature. Each square box indicates a deep neural network and arrows shows the average messages from neighbors. }
\label{fig:message-passing}
\end{figure*}

Conceptually speaking, the idea of GCN is closely related to neighborhood-aggregation encoder algorithm \cite{aggregation_alg}. However, instead of aggregating information from neighbors, the intuition behind GCNs is to view graphs from the perspective of \emph{message passing} algorithm between nodes \cite{messageGNN}. An illustration of message passing is depicted in Fig. \ref{fig:message-passing}.
In the GCN framework, every node is initialized with an embedding which is the same at its feature vector. At each layer of the GCN algorithm, nodes take the average of neighbor messages and apply a neural network on that as follows
\begin{equation}\label{GNN_expression}
\begin{aligned} 
{\bf h}_v^{(t)}=
\sigma({\bf W}_1^{(t)}{\bf h}_{v}^{(t-1)}+
{\bf W}_2^{(t)}\sum_{u:u\in {\mathcal N}(v)}{{\bf h}_{u}^{(t-1)}}),
\end{aligned}
\end{equation}
where ${\bf h}_{v}^{(t)}$, ${\bf W}_1^{(t)}$, ${\bf W}_2^{(t)}$, and $\sigma$ respectively represent the hidden vector of node $v$ at layer $t$, weight matrix at layer $t$ for self node, weight matrix at layer $t$ for neighboring nodes, and a non-linear function (e.g. ReLU). Furthermore, ${\mathcal N}(v)$ indicates the neighbors of node $v$. 
After $K$-layers of neighborhood aggregation, we get output embedding for each node. We can use these embeddings along with any
loss function and running stochastic gradient
descent to train GCN model parameters.


The above scheme is suitable for undirected graphs where edges show reciprocal relationship between ending nodes.

\subsection{Overview of GCNs for Directed Graphs}
In real applications such as social media where relations between nodes are not reciprocal, using regular GCN-based schemes might not be useful. An alternative for this type of situations is to utilize schemes deisgned for directed graphs such as EDGNN \cite{edgnn} where incoming and outgoing edges are treated differently in order to capture nonreciprocal relationship between nodes. In other words, EDGNN considers different weights for outgoing and incoming edges in addition to weights set for neighboring nodes. In particular, the embedding of node $v$ would be as follows:
\begin{equation}\label{EDGNN_expression}
\begin{aligned} 
{\bf h}_{n,v}^{(t)}&=
\sigma({\bf W}_1^{(t)}{\bf h}_{n,v}^{(t-1)}+
{\bf W}_2^{(t)}\sum_{u:u\in {\mathcal N}(v)}{{\bf h}_{n,u}^{(t-1)}}
\\&+
{\bf W}_3^{(t)}\sum_{u:e_{u,v}\in E_{Task}}{{\bf h}_{e,(u,v)}^{(t-1)}}
\\&+{\bf W}_4^{(t)}\sum_{u:e_{v,u}\in E_{Task}}{{\bf h}_{e,(v,u)}^{(t-1)}}
),
\end{aligned}
\end{equation}
where ${\bf W}_1^{(t)}$, ${\bf W}_2^{(t)}$, ${\bf W}_3^{(t)}$, and ${\bf W}_4^{(t)}$ represent weight matrices of layer $t$ for embedding of self node, neighboring nodes, incoming edges, and outgoing edges, respectively. Moreover, ${\bf h}_{n,v}^{(t)}$ and ${\bf h}_{e,(u,v)}^{(t)}$ respectively denote embedding of node $v$ and the embedding of the edge from node $u$ to node $v$ at layer $t$.  

\subsection{Proposed Input Graph}\label{implementation}
In order to train an EDGNN-based model, we need to carefully design the input graph components, namely adjacency matrix, nodes' features, edges' features, and labels. It should be noted that our scheme is not tailored to a particular criterion as we will show later that it can learn from two scheduling schemes with different objectives. Our deigned input graph can be fed into the EDGNN and the model will be trained according to labels generated from a given scheduling scheme. We next explain how we carefully design the input graph.

\noindent{\bf Designed Input Graph}: We start from the original task graph and consider the same set of nodes and edges for our input graph as the task graph. In other words,  by representing the input graph as $G_{input}:=(V_{input},E_{input})$, we have $V_{input}:=V_{Task}$ and $E_{input}:=E_{Task}$. The crucial part for having an efficacious GCN-based scheduler has to do with carefully designing the features of nodes and edges as well as the labels. 

$\bullet$ The feature of node $T_i$, $\forall T_i\in V_{input}$, is denoted by ${\bf x}_{{\bf n},i}$ and it has the following $N_C$-dimension features:
\begin{equation}\label{node_feature}
\begin{aligned} 
{\bf x}_{{\bf n},i} :=\Big(\frac{p_{i}}{e_1},\frac{p_{i}}{e_2},\cdots,\frac{p_{i}}{e_{N_C}}\Big)^T \in \mathbb{R}^{N_C}.\nonumber
\end{aligned}
\end{equation}
The intuition behind ${\bf x}_{{\bf n},i}$ is that these features represent the required computational time of task $T_i$ across all compute machines.

\begin{figure}[h]
\centering
\includegraphics[scale=.45,trim= 70mm 80mm 90mm 5mm,clip=true]{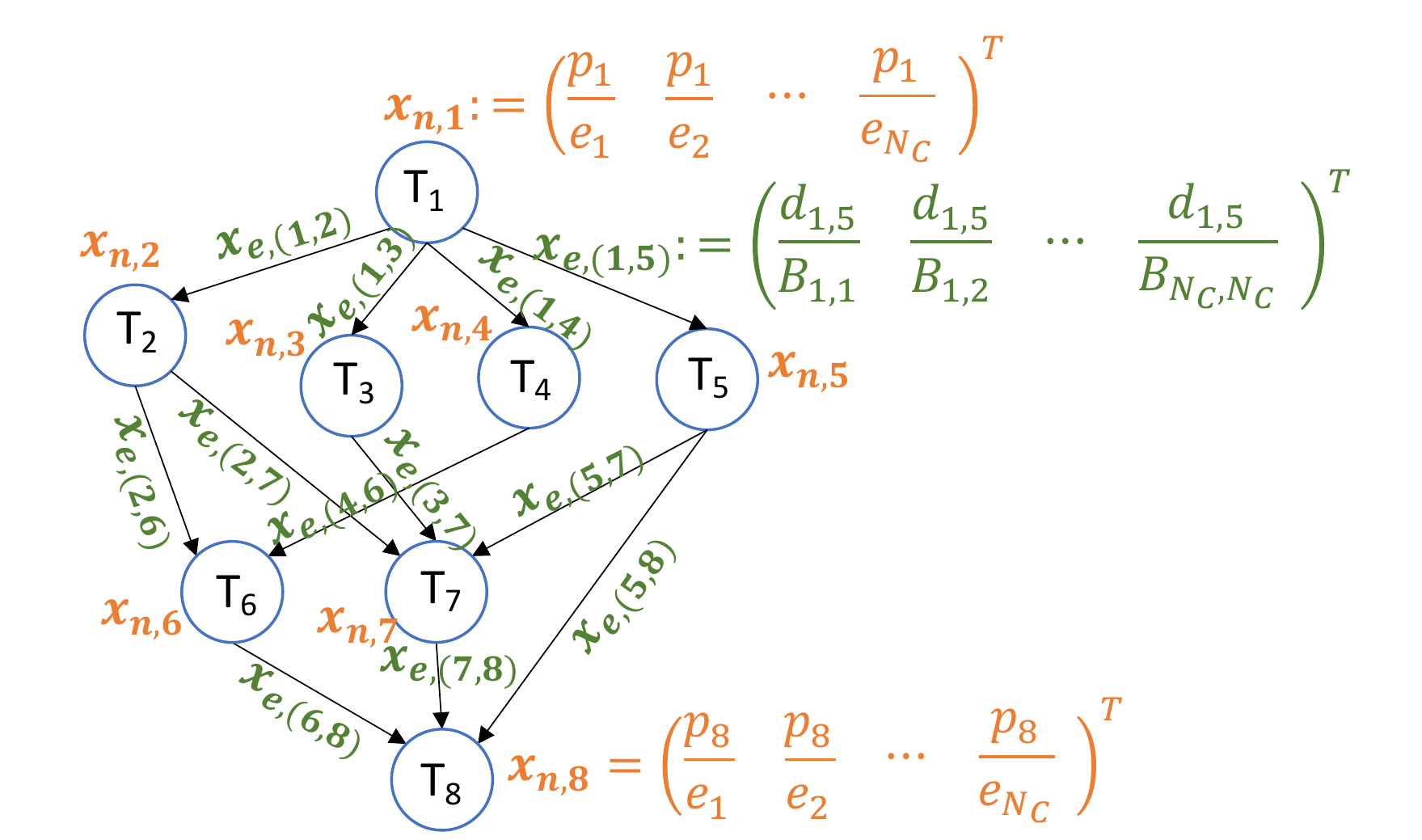}
\caption{An illustration of designed input graph fed into the EDGNN model for the example of the task graph shown in Fig. \ref{fig:task_graph}. Node and edge features are represented with brown and green colors, respectively.}
\label{fig:GNN}
\end{figure}

$\bullet$ The feature of edge $e_{u,v}$, $\forall e_{u,v}\in E_{input}$, is denoted by ${\bf x}_{{\bf e},(u,v)}$ and it has the following $N_C^2$-dimension features:
\begin{equation}\label{edge_feature}
\begin{aligned} 
{\bf x}_{{\bf e},(u,v)}:=\Big(\frac{d_{u,v}}{B_{1,1}},\frac{d_{u,v}}{B_{1,2}},\cdots,\frac{d_{u,v}}{B_{N_C,N_C}}\Big)^T \in \mathbb{R}^{N_C^2}.\nonumber
\end{aligned}
\end{equation}
The intuition behind ${\bf x}_{{\bf e},(u,v)}$ is that these features represent the required time for transferring the result of executing task $T_u$ to the following task $T_v$ across all possible pair-wise compute machines. An illustration of our designed input graph for the task graph of Fig. \ref{fig:task_graph} is depicted in Fig. \ref{fig:GNN}.

\textbf{Objective-Dependent Labeling:}
Based on what task scheduler our method should learn from (which we refer to as the ``teacher" scheduler, namely, HEFT for makespan minimization and TP-HEFT for throughput maximization), we label all nodes as well as edges. Let us define $L_{{\bf n},v}$ and $L_{{\bf e},(u,v)}$ as labels of node $v$ and edge $e_{u,v}$, respectively. Regarding nodes' labeling, we consider the label of node $T_i$, $\forall i$, as the index of compute node that the teacher algorithm assigns task $T_i$ to run on. 

Thus, for makespan minimization, we have that:
$L_{{\bf n},v}=m_{HEFT}(v)$ $\forall v\in V_{Input}$ where $m_{HEFT}(.)$ is the mapper function of HEFT algorithm.

And for throuhgput maximization, we have that:

$L_{{\bf n},v}=m_{TP-HEFT}(v)$ $\forall v\in V_{Input}$ where $m_{TP-HEFT}(.)$ is the mapper function of the TP-HEFT algorithm.

Finally, we label each edge according to the label of the ending vertex it is from. In other words, $L_{{\bf e},(u,v)}=L_{{\bf n},u}$ $\forall u,v$ such that $e_{u,v}\in E_{input}$. We should note that this edge-labeling is crucial in enforcing the model to learn to label out-going edges of a node with same label as its corresponding node's label.

\subsection{Implementation}\label{implementation}
As far as the model parameters are concerned, we consider a 4-layer EDGCN with 128 nodes per layer with $ReLU$ activation function and $0$ dropout. Since we suppose nodes and edges have features, we let both nodes and edges to be embedded. 

For training our model, we need to create a sufficiently large graph. However, since the HEFT and TP-HEFT algorithms are extremely slow in performing task-scheduling for large-scale task graphs, obtaining labels (i.e. determining the machine each task need to be executed on) for a single large graph is cumbersome. Therefore, we create a large graph $G_{union}$ by taking the union of disjoint medium-size graphs $G_{i}$'s (i.e., $G_{union}:=\{G_{i}\}_{i=1}^{N_g}$) such that HEFT and TP-HEFT can handle scheduling tasks over each of them. Regarding splitting dataset, we consider $60\%$, $20\%$, and $20\%$ of the disjoint graphs for training, validation, and test, respectively.  This allows us to apply the teacher algorithms to train our GCNScheduler for even larger graphs than the teacher algorithms themselves can handle efficiently. 

\section{Experimental Results}\label{results}
In this section, we evaluate the performance of our proposed GCNScheduler in terms of two criteria, namely the makespan minimization and the throughput maximization, for various task graphs (medium-scale and large-scale task graphs as well as the task graphs of three real perception applications considered in \cite{Odessa}). For each criterion, we measure the performance of GCNScheduler as well as the time it takes to assign tasks and compare these values with the corresponding values of our benchmarks (i.e. HEFT/TP-HEFT and the random task-scheduler). We evaluate all schemes by running them on our local cluster which has 16 CPUs (with 8 cores and 2 threads per core) of Intel(R) Xeon(R) E5-2620 v4 @ 2.10GHz.

As far as network settings are concerned, the computation amount of tasks, the execution speed of compute machines, and the communication bandwidth are drawn randomly from uniform distributions. 
For simplicity, we assume each task produces the same amount of data after being executed. 
Regarding task graphs, we generate random DAGs in two different ways: $1)$ establishing an edge between any two tasks with a given probability (which we call \emph{edge probability (EP)}), then pruning\footnote{Only keep edges between nodes $u$ and $v$ if $u<v$.} them such that it forms a DAG, $2)$ specifying the width and depth of the graph, then randomly selecting successive tasks for each task.

\subsection{Makespan Minimization}
For training the model, since the teacher scheduler, i.e. the HEFT algorithm~\cite{HEFT}, is extremely slow in generating labels for large task graphs\footnote{We observed that HEFT is incapable of conducting task-assignment running on a commodity PC when the task graph become large, on the order of 100 nodes, see table 1 for the increasing trend in compute time with task graph size.}, we create sufficient number (approximately 400) of random medium-size task graphs (i.e. each has $\leq$50 nodes with either an $EP$=0.25 or a width and depth of 5 and 10, respectively) and label tasks for each of these medium-size task graphs according to the HEFT algorithm~\cite{HEFT}. By doing so, we create a single large-scale graph which is the union of disjoint medium-size graphs. On the significance of our proposed scheme, it only takes $<15$ seconds to train our model with such a large graph using just CPU's (no GPU's or other specialized hardware). 
After training the model, we consider both medium-scale and large-scale task graphs as input samples. Then our model labels the tasks and determines what machine will execute each task. We next measure the performance of GCNScheduler over medium-scale and large-scale task graphs as well as the task graphs of the three real perception applications provided in~\cite{Odessa}. 

\subsubsection{Medium-scale task graphs} \label{sec:medium-HEFT}
\begin{figure}[h]
\centering
\includegraphics[scale=.6]{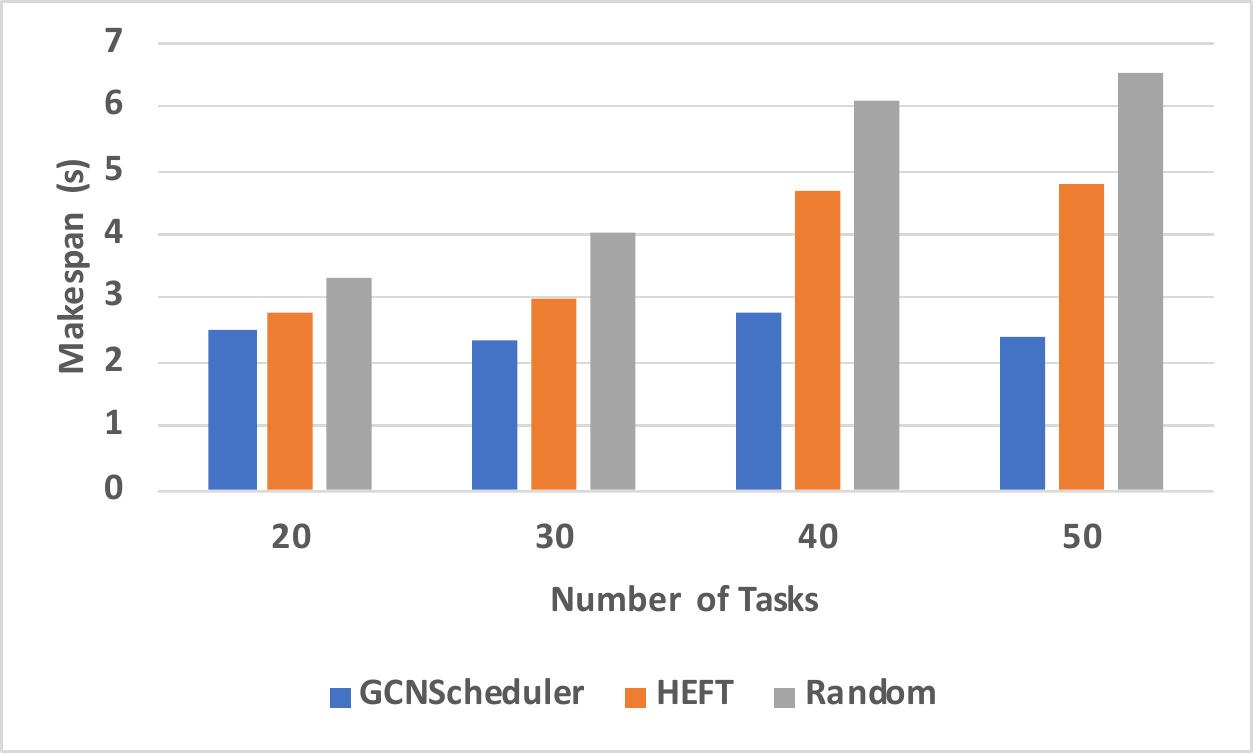}
\caption{
Makespan of GCNScheduler, HEFT \cite{HEFT}, and the random scheduler in small settings for different number of tasks with $EP=0.25$.
}
\label{fig:Makespan_medium}
\end{figure}

Fig. \ref{fig:Makespan_medium} shows the average makespan of GCNScheduler (with the makespan-minimization objective) compared to HEFT~\cite{HEFT} and the random task-scheduler for medium-size task graphs with different number of tasks. Our model significantly outperforms the random task-scheduler and considerably improves the makespan compared to HEFT, specially as number of tasks increases. 
The accuracy of GCNScheduler is around $\%75$ with respect to replicating the schedules produced by HEFT; however, we should note that ultimately the makespan of executing all tasks is more important rather than the accuracy of labeling tasks, and in this regard GCNScheduler does better than HEFT. To gain some intuitive understanding for why GCNSchduler outperforms HEFT in terms of makespan, we manually examined and observed the schedules produced by GCNScheduler and HEFT for many scenarios. We found that the HEFT is sometimes unable to prevent assigning tasks to machines with poor communication bandwidth while GCNScheduler is able to learn to do so more consistently. We believe this is due to the carefully-designed features of edges in the input to GCNScheduler which explicitly take communication bandwidth between machines into account.   
 
The time taken to assign tasks to compute machines for both GCNScheduler and HEFT~\cite{HEFT} is presented in Table \ref{table:task_assignment_time_small}. As one can easily see, our GCNScheduler outperforms HEFT by 3-7 orders of magnitude. This clearly shows GCNScheduler is a game-changer for task scheduling.   
\begin{table}[h]
\centering
\caption{Time taken (in seconds) by GCNScheduler and HEFT to perform scheduling for medium-scale task graphs with different number of tasks.}
\label{table:task_assignment_time_small}
\begin{tabular}{|c|cccc|}
\hline
& &   $~~~~~~~~~~N_T$ & &\\
& 20 & 30 & 40 & 50\\
\hline
\text{GCNSch.} & 0.0026&  0.0027 & 0.0029 & 0.0037\\
\text{HEFT} & 0.6764 & 9.4330 & 117.35 & 1552.0\\
\hline
\end{tabular}
\end{table}

\begin{figure*}[h]
\centering
\includegraphics[scale=0.5]{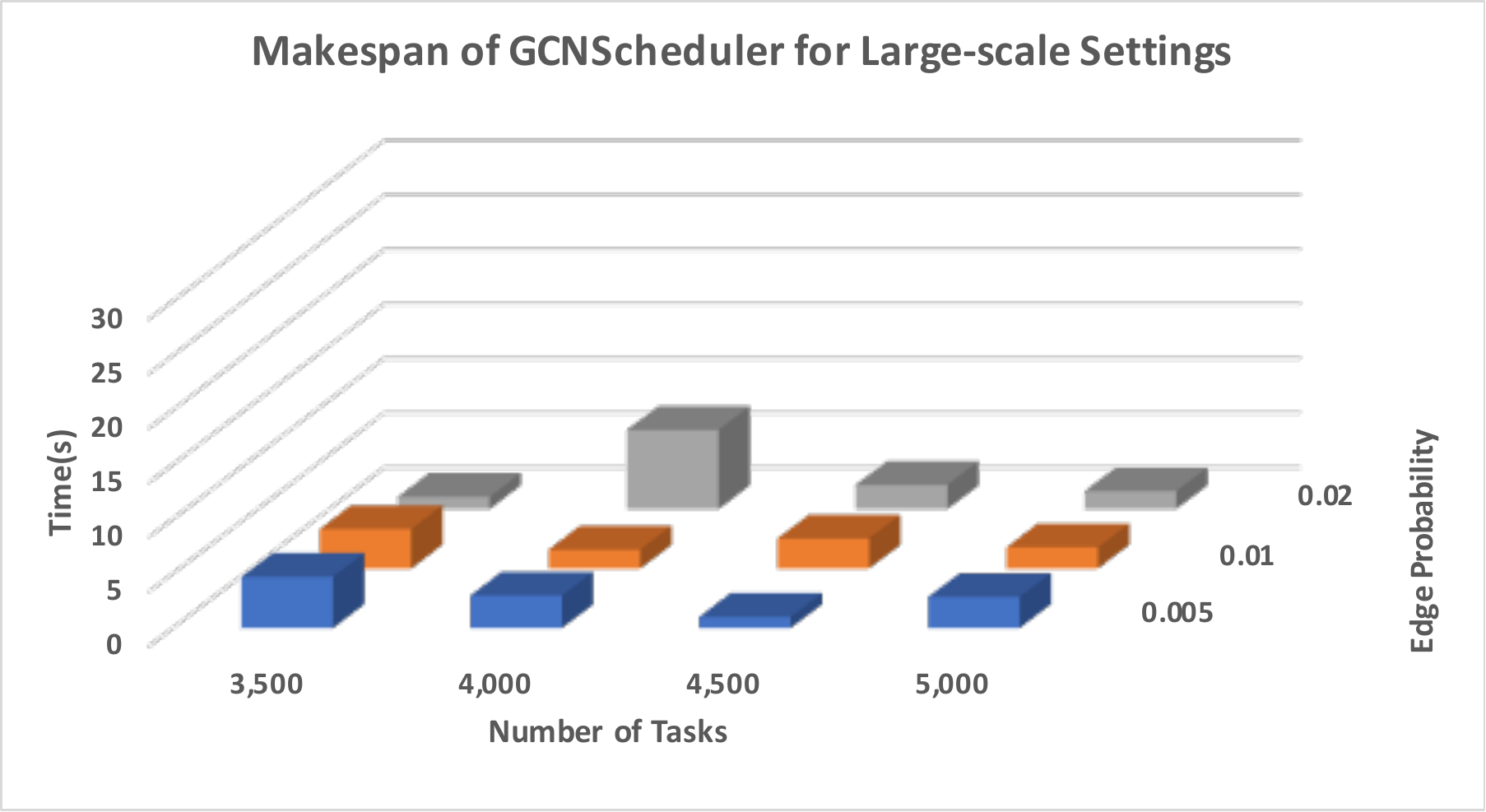}
\includegraphics[scale=0.5]{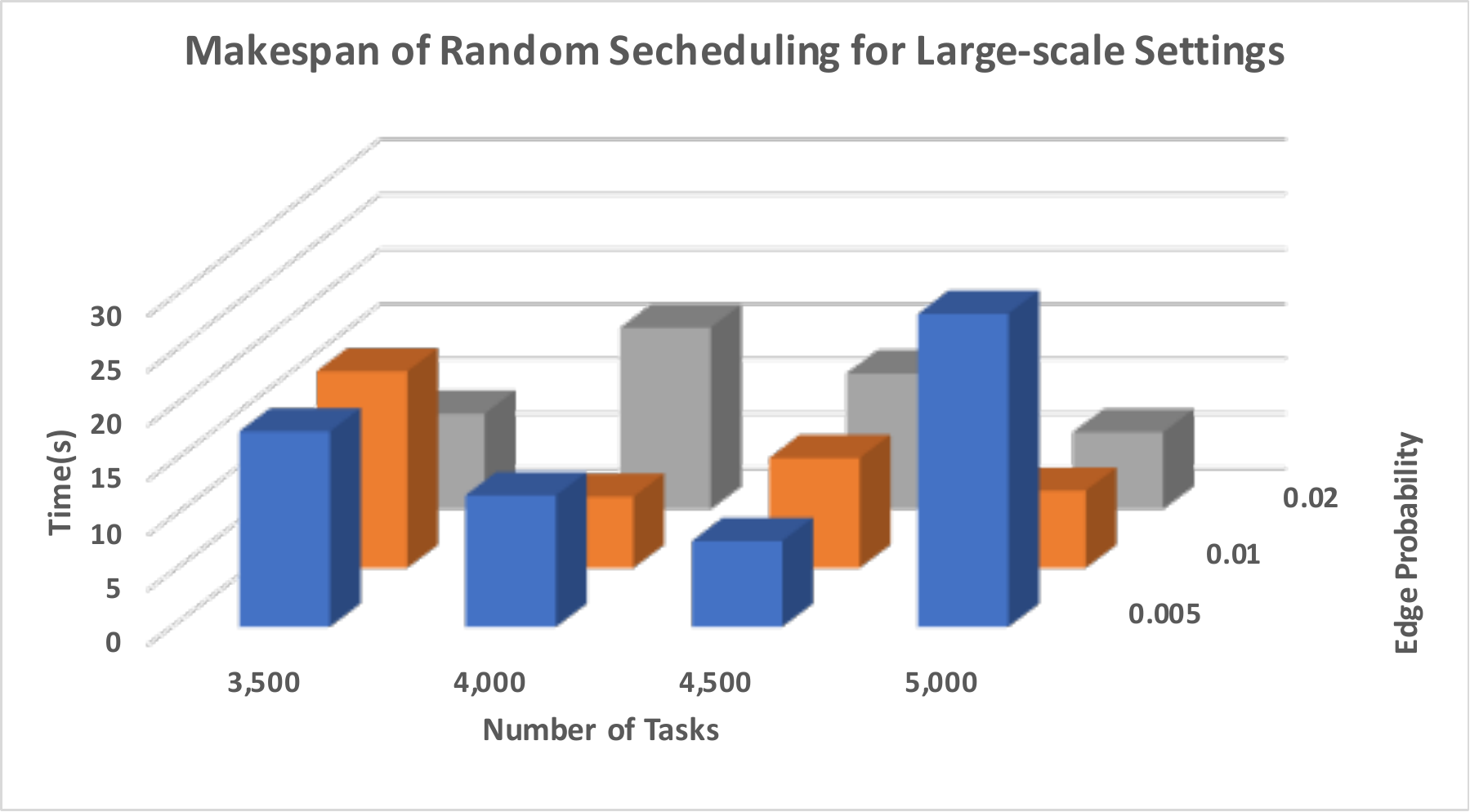}
\caption{Makespan of GCNScheduler and the random scheduler for large-scale task graphs with different number of tasks and different $EP$.}
\label{fig:Makespan_large}
\end{figure*}

\subsubsection{Large-scale task graphs}
In addition to the promising results on the medium-size settings, we now focus on large-scale task graphs where HEFT algorithm \cite{HEFT} is extremely slow to operate; hence we compare the performance of our GCNScheduler with only possible benchmark which is the random task-scheduler. Fig. \ref{fig:Makespan_large} shows the average makespan of our proposed GCNScheuler (top plot) and the random task-scheduler (bottom plot) in large-scale settings where number of tasks varies from 3,500 to 5,000 and the edge probability (i.e. $EP$) takes 0.005, 0.01, and 0.02. One can easily observe that our proposed GCNScheduler significantly reduces makespan by a factor of 8 (for larger $EP$). The intuition behind the significant gain for larger $EP$ (i.e. node's degrees are larger) is that some tasks may require more predecessor tasks to be executed in advance (because of having larger nodes' degree), hence randomly assigning tasks may potentially assign one of the predecessor task to a machine with poor computing power or communication bandwidth, resulting in a larger average makespan. However, GCNScheuler efficiently exploits inter-task dependencies as well as network settings information (i.e. execution speed of machines and communication bandwidth across machines) through carefully-designed node and edge features; therefore it leads to a remarkably lower makespan.       
\begin{figure}[h]
\centering
\includegraphics[scale=.55]{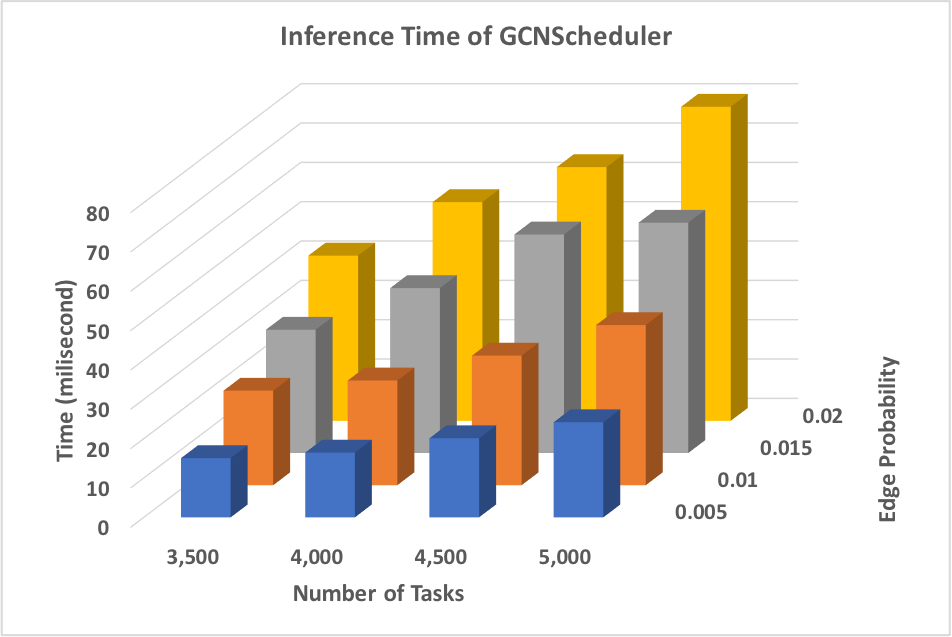}
\caption{
Inference time of our GCNScheduler for large-scale task graphs with different number of tasks and different $EP$.
}
\label{fig:inference}
\end{figure}


Finally, Fig. \ref{fig:inference} illustrates the inference time (i.e. time taken to assigning tasks to compute machines) of our proposed GCNScheduler for different number of tasks and different $EP$. Our GCNScheduler takes $<80$ milliseconds to obtain labels for each of these large-scale task graphs. This clearly shows the great advantage of our proposed scheme which makes it an ideal alternative to state-of-the-art scheduling schemes which fail to efficiently operate over complicated jobs each of which may have thousands of tasks with any inter-task dependencies. 

\begin{figure}[h]
\centering
\includegraphics[scale=.7]{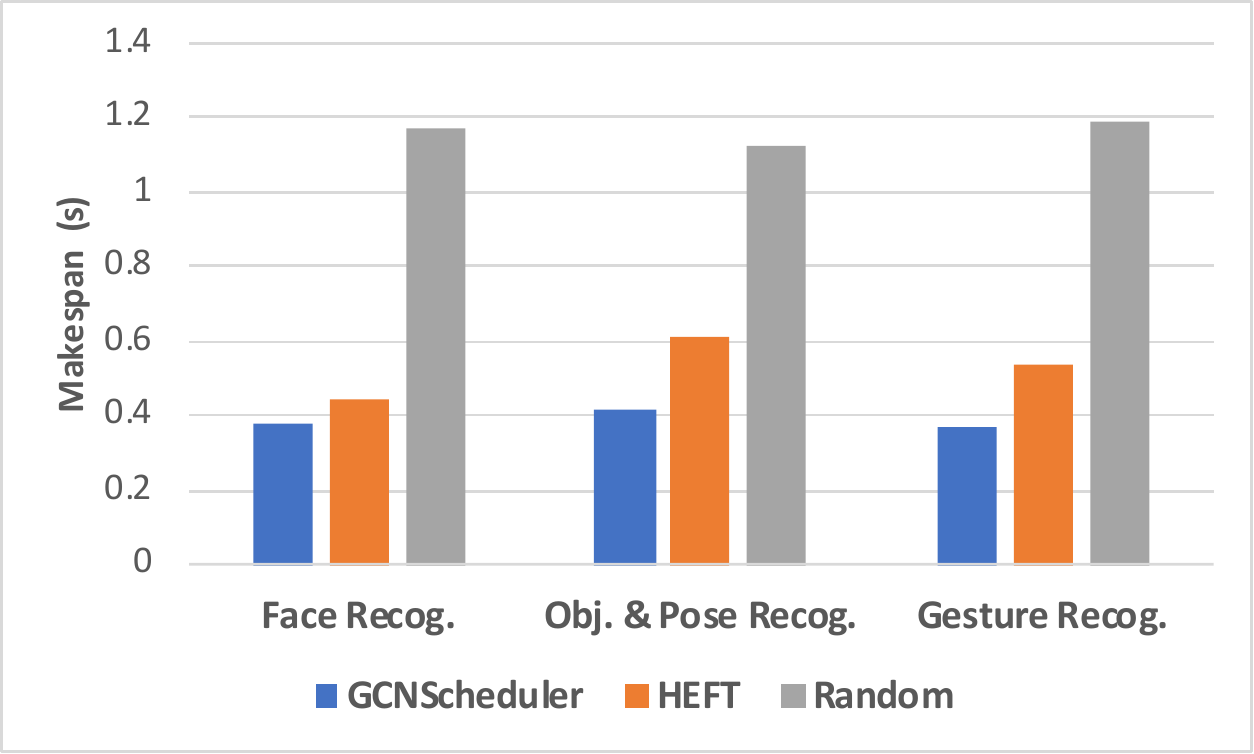}
\caption{Makespan of GCNScheduler (with the makespan-minimization objective), HEFT, and the random task-scheduler for the three real perception applications considered in \cite{Odessa}.
}
\label{fig:Odessa_HEFT}
\end{figure}

\begin{table}
\[\begin{array}{|c|ccc|}
\hline
  & \text{\shortstack{Face Recog.}} & \text{\shortstack{Pose Recog.}} & \text{\shortstack{Gesture Recog.}}\\
\hline
\text{GCNSch.} & 0.460 &  0.524 & 0.621 \\
\text{HEFT} & 227.7 & 1484 & 443.0 \\
\hline
\end{array}\]
\caption{Time taken (in {\bf milliseconds}) by GCNScheduler and HEFT \cite{HEFT} to perform scheduling for the task graph of the three real perception applications considered in \cite{Odessa}.}
\label{table:Odessa_HEFT}
\end{table}

\subsubsection{Real Application Benchmarks}\label{sec:perception:HEFT}
\begin{figure*}
\centering
\includegraphics[scale=.48,trim=15mm 3mm 0mm 20mm,clip=true]{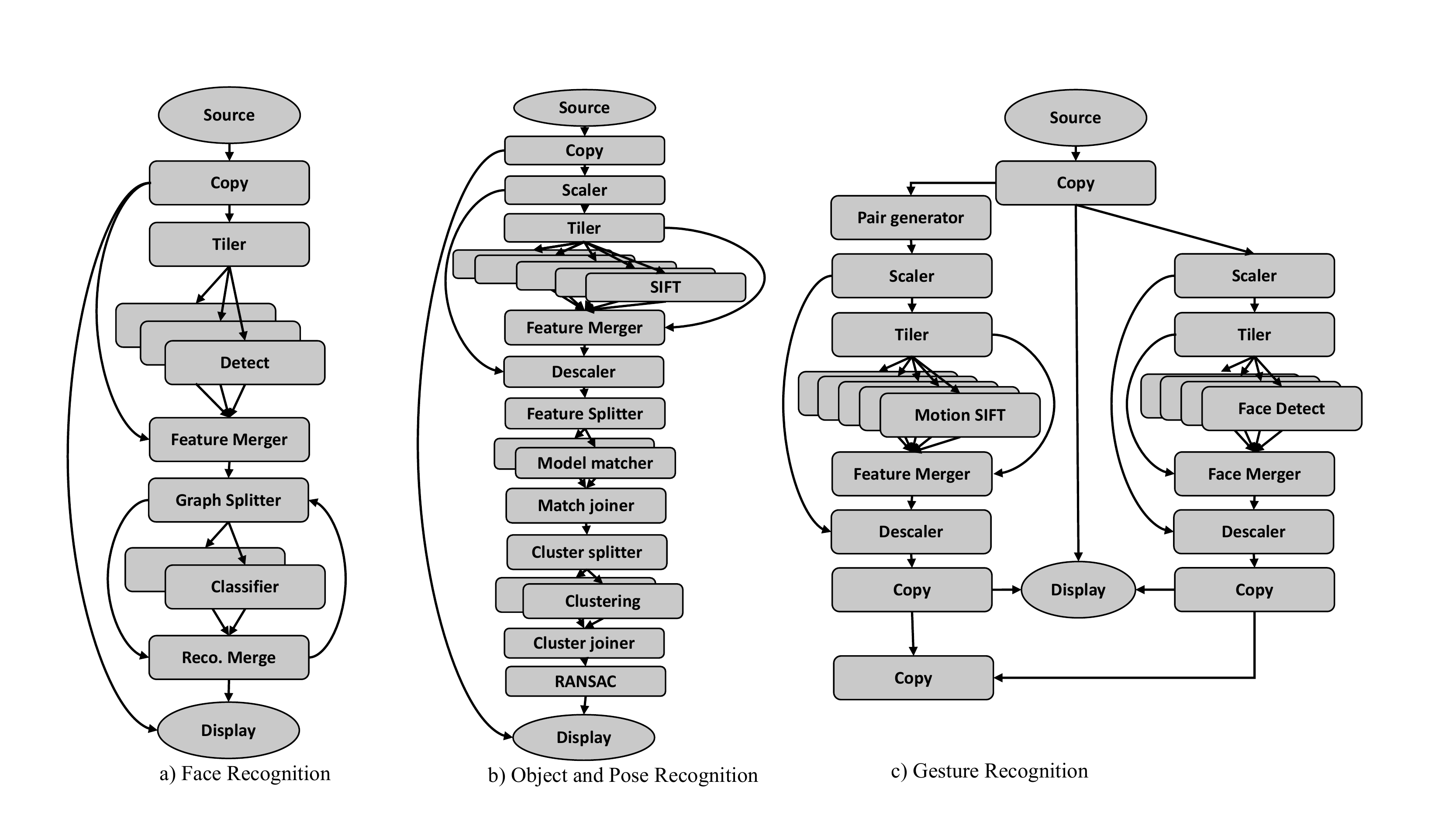}
\caption{
Task graphs of the three perception application considered in \cite{Odessa}.
}
\label{fig:perception_task_graphs}
\end{figure*}

To show the significance of our proposed scheme on real applications, we consider the three real perception applications provided in \cite{Odessa}, namely the face recognition, the object-and-pose recognition, and the gesture recognition with corresponding task graphs depicted in Fig. \ref{fig:perception_task_graphs} (Please see \cite{Odessa} for more detail regarding each task). 
We measure the makespan of each application by running GCNScheduler (with the makespan-minimization objective) over the carefully-designed input graphs (obtained from original task graphs). Fig. \ref{fig:Odessa_HEFT} illustrates the makespan of GCNScheduler (with the makespan-minimization objective) against HEFT \cite{HEFT} and the random task-scheduler for the three perception applications. While our GCNScheduler leads to a slightly better makespan performance compared to HEFT \cite{HEFT}, it significantly (by 3-4 orders of magnitude) reduces the time taken to perform scheduling as it is shown in Table \ref{table:Odessa_HEFT}. This remarkable time reduction demonstrates the importance of our scheme to other applications (in business or security) used/built upon these perception applications.

\subsection{Throughput Maximization}
For the purpose of maximizing the throughput, we use the TP-HEFT algorithm \cite{HEFT-TP} as the teacher scheduler for training our GCNScheduler. 
Since the TP-HEFT scheduler~\cite{HEFT-TP}, similar to the HEFT algorithm~\cite{HEFT}, excessively becomes slow in generating labels for large task graphs, we create sufficient number of random medium-size task graphs (i.e. each of which has around 40 tasks with the width and depth of 5 and 8, respectively) and label tasks according to the TP-HEFT algorithm \cite{HEFT-TP}. We then build a single large-scale graph, which is the union of disjoint medium-size graphs, and train GCNScheduler with the throughput-maximization objective. Similarly, we test our GCNScheduler over medium-scale and large-scale task graphs as well as the task graph of the three perception applications.

\subsubsection{Medium-scale task graphs}\label{medium-TP}
Table \ref{table:task_assignment_time_small_TP} shows the throughput of GCNScheduler (with the throughput-maximization objective) compared to TP-HEFT \cite{HEFT-TP} and the random task-schedulers for medium-size task graphs with different number of tasks. GCNScheduler leads to slightly higher throughput compared to TP-HEFT \cite{HEFT-TP} scheduler, while it significantly outperforms random task-scheduler. 
Table \ref{table:task_assignment_time_small_TIME_TP} also shows the time taken to schedule tasks. Moreover, the accuracy of our model is around $\%95$. 

\begin{table}
\[\begin{tabular}{|c|cccc|}
\hline
& &   $~~~~~~~~~~N_T$ & &\\
& 100 & 200 & 300 & 400\\
\hline
\text{GCNSch.} & 3.1254 &  3.1251 & 3.1261 & 3.1185\\
\text{TP-HEFT} & 2.1731 & 2.1690 & 1.8034 & 2.0046\\
\text{Random} & 0.0193 & 0.01801 & 0.0174 & 0.0177\\
\hline
\end{tabular}\]
\caption{Throughput of GCNScheduler (with the throughput-maximization objective), Throughput(TP)-HEFT algorithm, and the random task-scheduler for medium-size task graphs with different number of tasks.}
\label{table:task_assignment_time_small_TP}
\end{table}

\begin{table}
\[\begin{tabular}{|c|cccc|}
\hline
& &   $~~~~~~~~~~N_T$ & &\\
& 100 & 200 & 300 & 400\\
\hline
\text{GCNSch.} & 0.0033&  0.0049 & 0.0071 & 0.0078\\
\text{TP-HEFT} & 6.9235 & 27.229 & 70.940 & 115.221\\
\hline
\end{tabular}\]
\caption{Time taken (in seconds) by GCNScheduler (with the throughput-maximization objective) and TP-HEFT to schedule for medium-size task graphs with different number of tasks.}
\label{table:task_assignment_time_small_TIME_TP}
\end{table}

\subsubsection{Large-scale task graphs}
Since TP-HEFT \cite{HEFT-TP} and other existing scheduling schemes are extremely slow for very large task graphs (e.g. task graphs with few thousands tasks), we only compare the throughput of SCNScheduler against the random task-scheduler, as shown in Table \ref{table:TP_large}. Further, Table \ref{table:time_TP_large} shows the time taken for assigning tasks to compute nodes for large-scale task graphs. One can easily see GCNScheduler (with the throughput-maximization objective) is remarkably fast while handling large-scale task graphs.
\begin{table}
\begin{tabular}{|c|cccc|}
\hline
&&$~~~~~~~~N_T$&&\\
 & 3,500 & 4,000 & 4,500 & 5,000\\
\hline
\text{GCNSch.} & 2.9737 &  2.9731 & 2.9733 & 2.9734\\
\text{Random} & 0.6344 & 0.6336 & 0.6332 & 0.6331\\
\hline
\end{tabular}
\caption{Throughput of GCNScheduler (with the throughput-maximization objective) and the random task-scheduler for large-scale task graphs.}
\label{table:TP_large}
\end{table}

\begin{table}
\begin{tabular}{|c|cccc|}
\hline
&&$~~~~~~~~N_T$&&\\
 & 3,500 & 4,000 & 4,500 & 5,000\\
\hline
\text{GCNSch.} & 66.050 &  74.978 & 83.817 & 87.388\\
\hline
\end{tabular}
\caption{Time taken (in {\bf milliseconds}) by GCNScheduler (with throughput objective) to schedule.}
\label{table:time_TP_large}
\end{table}

\subsubsection{Real Application Benchmarks}\label{sec:perception:TP}
\begin{figure}
\centering
\includegraphics[scale=.7]{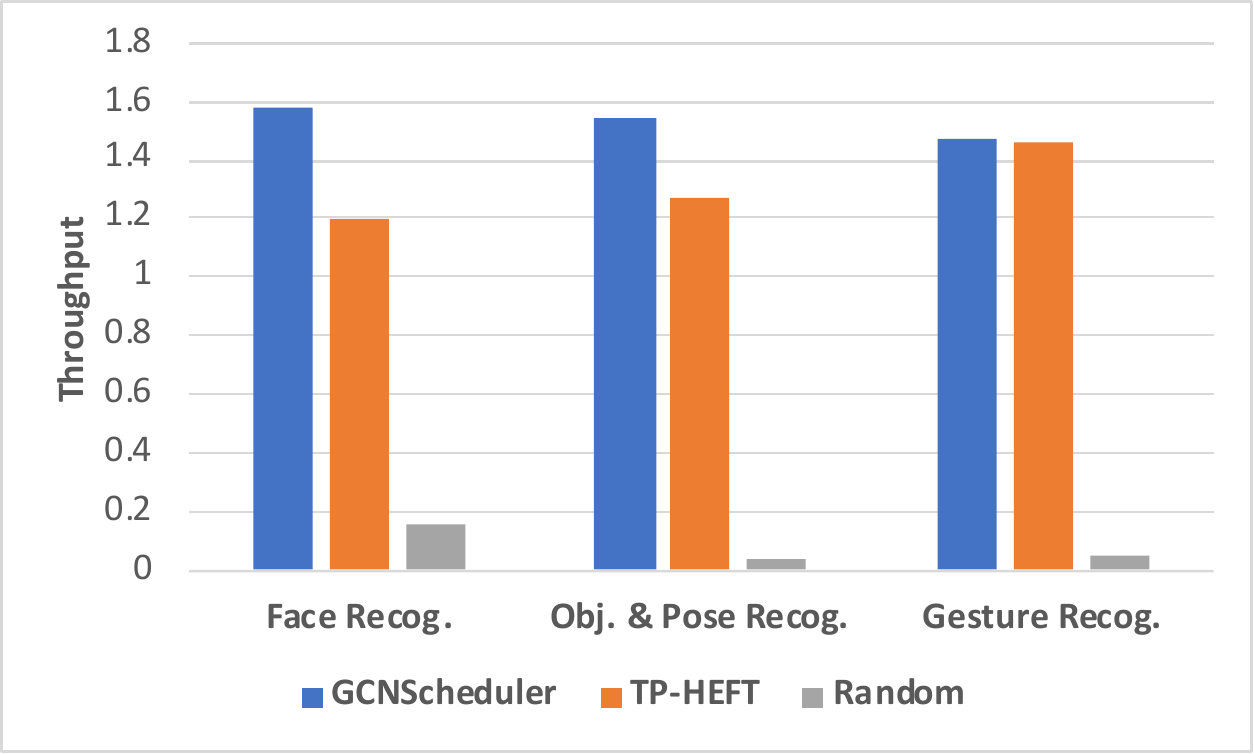}
\caption{
Throughput of GCNScheduler (with throughput objective),  TP-HEFT, and random scheduler for the task graph of the three real perception applications.
}
\label{fig:Odessa_TP}
\end{figure}

We now evaluate the throughput of our GCNScheduler, given the task graph of the three real perception applications provided in \cite{Odessa}, namely the face recognition, the object-and-pose recognition, and the gesture recognition. In particular, we run our trained GCNScheduler (with the throughput-maximization objective) over the carefully-designed input graphs and measure the throughput for each application. Fig. \ref{fig:Odessa_TP} shows the throughput of our GCNScheduler (with the throughput-maximization objective) compared to TP-HEFT \cite{HEFT-TP} and the random task-scheduler for the three perception applications. While our GCNScheduler leads to a marginally better throughput performance compared to TP-HEFT scheduler \cite{HEFT-TP}, it significantly (2-3 orders of magnitude) reduces the time taken to perform task-assignment as it is shown in Table \ref{table:Odessa_TP}.

\begin{table}
\caption{Time taken (in {\bf milliseconds}) by GCNScheduler (with the throughput-maximization objective) and TP-HEFT to perform scheduling for the task graph of the three real perception applications.}
\begin{tabular}{|c|ccc|}
\hline
& \text{\shortstack{Face Recog.}} & \text{\shortstack{Pose Recog.}} & \text{\shortstack{Gesture Recog.}}\\
\hline
\text{GCNSch.} & 0.488 &  0.511 & 0.560 \\
\text{TP-HEFT} & 87.34 & 257.8 & 290.1 \\
\hline
\end{tabular}
\label{table:Odessa_TP}
\end{table}

\section{Conclusion}
We proposed GCNScheduler, a scalable and fast task-scheduling scheme which can perform scheduling according to different objectives (such as  minimizing the makespan or maximizing the throughout). By evaluating our scheme against benchmarks through simulations, we show that not only can our scheme easily handle large-scale settings where existing scheduling schemes are unable to do, but also it can lead to a better performance with significant lower required time (i.e., several orders of magnitude faster) to perform scheduling. 
As our future direction, we aim at investigating the performance of our proposed GCNScheduler with respect to other objectives with other teacher schedulers.

\subsubsection*{Acknowledgments}
This material is based upon work supported in part by Defense Advanced Research Projects Agency (DARPA) under Contract No. HR001117C0053 and by the Army Research Laboratory under Cooperative Agreement W911NF-17-2-0196. Any views, opinions, and/or findings expressed are those of the author(s) and should not be interpreted as representing the official views or policies of the Department of Defense or the U.S. Government.

\bibliography{example_paper}
\bibliographystyle{iclr2022_conference}


\end{document}